\documentclass[prb,twocolumn,showpacs,preprintnumbers,amsmath,amssymb]{revtex4}
\usepackage{graphicx}% Include figure files
\usepackage{dcolumn}% Align table columns on decimal point
\usepackage{bm}% bold math
\newcommand{\cminv}{\mbox{cm$^{-1}$}}

\begin{document}

\title{\textbf{Large nonlinear absorption and refraction coefficients of
carbon nanotubes estimated from femtosecond z-scan measurements} }

\author{\textbf{N. Kamaraju, Sunil Kumar, and A. K. Sood\footnote{Corresponding author, email: asood@physics.iisc.ernet.in}}}
\affiliation{Department of Physics and CULA, Indian Institute of
Science, Bangalore 560012,India}
\author{Shekhar Guha}
\affiliation{Materials Directorate, WPAFB, Dayton, OH 45433, USA}
\date{\today}
\author{Srinivasan Krishnamurthy}
\affiliation{SRI International, Menlo Park, CA 94025, USA}
\author{C. N. R. Rao}
\affiliation{CPMU, Jawaharlal Nehru Centre for Advanced Scientific
Research,Bangalore 560064, India }
%\draft
\begin{abstract}

Nonlinear transmission of $80$ and $140$ femtosecond pulsed light
with $0.79~\mu m$ wavelength through single walled carbon
nanotubes suspended in water containing sodium dodecyl sulphate is
studied. Pulse-width independent saturation absorption and
negative cubic nonlinearity are observed, respectively, in open
and closed aperture z-scan experiments. The theoretical
expressions derived to analyze the z-dependent transmission in the
saturable limit require two photon absorption coefficient
$\beta_0\sim$ $1.4~cm/MW$ and a nonlinear index $\gamma \sim -5.5
\times10^{-11}~ cm^2/W$ to fit the data.

       ~~~~~~~~~~~~~~~~~~~~~~~~~~~~~~(Accepted and to appear in Applied Physics Letters)
\end{abstract}
%{*Corresponding author. Fax :+91-080-23602602; Email:asood@physics.iisc.ernet.in}


\maketitle Single walled carbon nanotubes (SWNTs) have been
studied for numerous applications including third order optical
nonlinearity. These applications include nanoelectronics, gas and
bio sensors, field emission displays, saturable absorbers for
passive optical regeneration, mode-locking and THz optical
switching.  The suspended carbon nanotubes show optical limiting
due to nonlinear scattering, micro-plasma formation and
sublimation in the nanosecond
regime.\cite{Chen,Liu,sood,Vivien,Sun} In the femtosecond ($fs$)
regime where heating does not play a role in the nonlinear
transmission, an enormously large third-order susceptibility
$(Im\chi^{(3)}\sim10^{-6}~esu)$ has been reported by resonantly
exciting at the first inter sub-band energy levels ($S_{11}$) of
semiconducting SWNTs.\cite{Maeda,achiba} The origin of this large
nonlinearity is assigned to the coherence effect, rather than the
incoherent or saturable absorption, because the measured
nonlinearity decreased rapidly when the wavelength is changed away
from the first band gap.\cite{Maeda} This interpretation for the
large nonlinearity is being debated.\cite{Sheng,Okamoto} Other
experiments\cite{Ajayan,Tatsuura} using fs pulses have reported a
mixed variation with wavelength-much smaller value\cite{Ajayan}
for Im$\chi^{(3)}\sim 10^{-10}$ esu at $1.55~\mu m$ and a fairly
large value\cite{Tatsuura} of $\sim 10^{-7}~esu$ at $1.33~\mu m$.
The origin of the nonlinearity had not been identified. In this
letter we report the results of closed aperture (CA) and open
aperture (OA) z-scan measurements carried out in suspension of
almost isolated SWNTs at a wavelength of $0.79~\mu m$ (energy
$\hbar \omega$ of $1.57~eV$). The OA z-scan shows saturated
absorption and CA z-scan reveals negative cubic nonlinear
refraction.  We have developed a theoretical model incorporating
saturated absorption along with nonlinear absorption and
refraction to derive the transmission in both the OA and CA
z-scans. The theoretical analysis of our results obtained with two
different pulses with full-width at half-maximum (FWHM) of $80~fs$
and $140~ fs$ clearly identifies two photon absorption (TPA) as
the source of nonlinearity and the TPA coefficient $\beta$ is
$\sim1.4 ~cm/MW$. This translates to a fairly large value of
\noindent $\sim 1\times 10^{-9}~esu$ for $Im\chi^{(3)}$
nonlinearity.

    A dispersion of SWNTs ($0.4~mg$) and $1\%$ of sodium dodecyl
sulphate (SDS) in water ($1~ml$) was sonicated for $5$ hours and
the resultant solution was found to be well dispersed. This
solution in $1~mm$ path length quartz cell was used in our
experiments. The SWNT sample used in our experiments is pristine
and contains two diameter distributions at $1.41~nm$ and $1.58~nm$
as confirmed by the radial breathing Raman modes at $160~\cminv$
and $177.7~\cminv$, shown in the inset of Fig. 1. \noindent

\begin{figure}[htbp]
\includegraphics[width=0.4\textwidth]{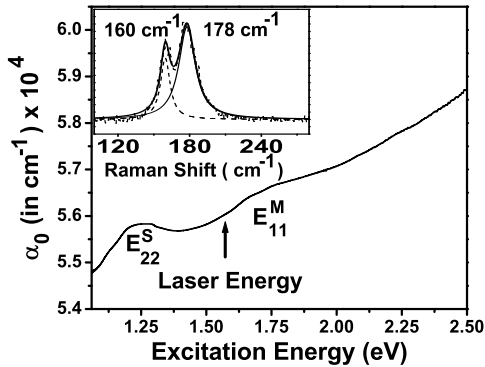}
\caption{~Optical absorption spectrum of SDS suspended Single
walled carbon nanotubes in water. $E_{22}^{S} (E_{11}^{M})$ are
the second (first) interband transition energy of semiconducting
(metallic) nanotubes. The inset is the Raman spectrum showing
radial breathing modes}
\end{figure}

 Nonlinear transmission studies were
carried out at $1.57~eV$ using Ti: Sapphire Regenerative
femtosecond amplifier (Spitfire, Spectra Physics). The chosen
photon energy is nearly resonant with the first interband
transition energy in metallic tubes, $M_{11}$ and off-resonant
with the second interband transition energy of the semiconducting
tubes $S_{22}$. From the absorption spectra of the dispersed
nanotubes shown in Fig. 1, we infer that the absorption
coefficient at this wavelength is about $5.6 \times 10^{4}
~\cminv$. The FWHM pulse width of the amplifier output was $50~fs$
at a repetition rate of $1~kHz$. Near the sample point, the pulse
from the amplifier was found to be broadened to $80~fs$. For the
experiments done with $140~fs$, we stretched pulse by adjusting
the compressor of the amplifier. Two Si-PIN diodes (one for the
signal ($B$) and the other for reference ($A$)) triggered at the
electronic clock output ($1~ kHz$) from the amplifier are used for
the data acquisition and the difference between $B$ and $A$ was
collected using a lock-in amplifier (SRS 830), averaged over 300
shots. This difference data was then converted into actual $B/A$
signal in a personal computer. The SWNT dispersion in $1~mm$ thick
cuvette was translated using a motorized translation stage (XPS
Motion controller, Newport) over the focal region. The intensity
of input beam was varied from $150~MW/cm^{2}$ to $6.2~GW/cm^{2}$.
In the OA z-scan, all the light was collected by using a
collection lens in front of the diode. The measured (and
normalized) transmission data as shown in Fig. 2(a) clearly
demonstrates the saturable absorption where the transmission is
enhanced at focus ($z=0$).  For CA z-scan, we kept an aperture of
diameter $3.6~mm$ in front of the diode $B$ and the measured
transmission (and normalized) data as shown in Fig. 2(b) indicates
photo-carrier induced reduction in the refractive index.

    The procedure to calculate the transmission in z-scan experiments
has been well described\cite{SB,Wearie,Chapple,Yin} in the
literature. For the optical limiting case, fairly accurate
solutions have been used to explain the z-scan
results.\cite{SB,Wearie,Chapple,Yin} However, for the saturation
absorption case, the solutions are obtained either qualitatively
\cite{Gao}or in the limit where the intensity is far smaller than
the saturation intensity.\cite{Yang} To quantify the basic
mechanisms responsible for the nonlinear absorption and
refraction, we accurately solve the rate equation, but
modified\cite{Samoc} for saturation absorption: \noindent

\noindent
\begin{equation}
    \frac {dI}{d\zeta} = -\frac{\alpha_0 I}{1+ \frac{I}{I_{s}}} -
    \beta_{0} I^{2}- \sigma_{0}\Delta N I
\end{equation}
\noindent where $\alpha_{0}$ is the one photon absorption
(including intrinsic free carrier absorption) coefficient,
$\beta_{0}$, is the fundamental TPA coefficient, and $\sigma_{0}$
is free carrier absorption (FCA) cross section. $I_{s}$ is the
parameter that characterizes the saturation absorption. The
intensity ($I$) at the radial position $r$, time $t$, the position
$\zeta$ in the sample, and the location of the sample $z$ is
denoted as $I(z,\zeta,r,t)$.  The generated photo-carrier density
$\Delta N$ depends on both $\alpha_{0}$ and $\beta_{0}$. In our
experiment, the maximum value of $I$ is $\sim$ $6~GW/cm^2$ and
$\alpha_{0}$ is $= 5.6 \times 10^4~\cminv$. Even with a large
value for $\beta_{0}$ $\sim 10^{-6}~cm/W$, the dominant source of
carrier generation is one photon absorption. Since the carrier
decay time is much longer than the pulse width $\tau_{0}$, Eq. 1
takes the form:
 \noindent
\begin{equation}
    \frac {dI}{d\zeta} = -\frac{\alpha_0 I}{1+ \frac{I}{I_{s}}} -
    \beta_{\rm eff} I^{2}.
\end{equation}
%where $\beta_{\rm eff} = \beta_0 + \frac {\sigma_0
%\alpha_0\tau_0}{\hbar\omega}$~~~~~~~~~~~~~~~~~~~~~~~~~~(2)

where \noindent \begin{equation}\beta_{\rm eff} = \beta_0 +
(\sigma_0 \alpha_0\tau_0 / \hbar\omega)
\end{equation}

The boundary condition required to solve Eq. 2 is the input
intensity which is assumed to be a Gaussian and
\begin{equation}
    I(z,0,r,t) = I_{0} \left[\frac{w_{0}}{w(z)}\right]^2
    exp\left[-\frac{2r^2}{w(z)^2}\right]exp\left[-\frac{t^2}{{\tau_0}^2}\right]
\end{equation}

\noindent $w_0$ is the beam waist at the focus,
$w(z)=w_0\sqrt{1+(z/z_0)^2}$ is the beam radius at $z$, $z_0=\pi
w^2_0/\lambda$ is the diffraction length of the beam, $\tau_0$ is
the half width at $e^{-1}$ of the maximum of the pulse, and
$\lambda$ is the wavelength. The intensity at the exit side of
sample, $I(z,L,r,t)$ is obtained from the analytical
solution\cite{Gradshteyn} to Eq. 2 and integrated over all $r$ and
$t$ to calculate the transmitted energy. The transmission in the
OA z-scan experiment, $T_{OA}(z)$ is simply the ratio of
transmitted energy to the incident energy.  The solution to Eq. 2
depends on two parameters $\beta_{\rm eff}$ and $I_s$, which can
be varied to fit the measured OA z-scan data.  Since several sets
of $\beta_{\rm eff}$ and $I_s$ can fit the data, we will choose
appropriate set that fits CA data as well.

    The transmission for CA case is more complicated as it requires
the phase of electric field~($E$) in addition to the intensity at
the exit surface.\cite{SB,Chapple,Yin} The phase at $\zeta=L$ is
different from that at $\zeta=0$ because of the change in
refractive index, $\Delta n$, caused by light absorption. To a
first order in $I$, the $\Delta n$ is simply $\gamma I$, where
$\gamma$ is the cubic nonlinear index.

    The cubic nonlinear refraction coefficient, $\gamma$, in general
has contribution from the photo-generated carriers, temperature
change, and bound electrons. Following the well established
procedure\cite{SB}, we calculate the electric field inside the
sample. For thin samples where photo-carrier generation is
uniform, the change in phase from incident to exit surface is
$\Delta\phi = k \gamma \int^{L}_{0}I(\zeta) d\zeta$, where k is
the wavenumber and $I(\zeta)$ is the solution of Eq. 2. For a
given r and t, the intensity at any $\zeta$ inside the sample is
obtained analytically, fitted to polynomial series in
$I(z,0,r,t)$, and substituted for $\Delta\phi$ to get,

\begin{eqnarray}
I(z,\zeta)&=&\sum_{n=0}^{6} a_{n}(\zeta) [I(z,0)]^{n+1};~~~~~ A_n
= \int^{L}_{0}a_{n}(\zeta) d\zeta \nonumber \\
\Delta\phi &=& k\gamma~
e^{-\frac{2r^2}{w(z)^2}-\frac{t^2}{{\tau_0}^2}}\sum_{n=0}^{6}
A_{n}I_0^{n+1}\left[\frac{w_{0}}{w(z)}\right]^{2n+2}
\end{eqnarray}

\noindent The electric field at the exit surface is then,

\begin{equation}
E(z,L,r,t)=\sqrt{I(z,L,0,0)}e^{-\left[\frac{r^2}{w(z)^2}+\frac{ikr^2}{2R(z)}+\frac{t^2}{{\tau_0}^2}-i\Delta\phi\right]}
\end{equation}

\noindent where $R(z)=z[1+z^2_{0}/z^2]$  is the radius of
curvature of the wave-front at $z$. As before\cite{SB},
$e^{i\Delta\phi}$ is expanded in infinite series and Gaussian
decomposition method\cite{Wearie} is used to obtain the field
pattern at the aperture which is at a distance d away from the
exit surface. In our calculations we found that the convergence is
achieved with first four terms in the expansion. We get normalized
z-scan transmittance, $T_{CA}(z)$ by integrating
$|E(z,L+d,r,t)|^2$ over all $t$ from $-\infty$ to $+\infty$, and
$r$ from $0$ to aperture radius $r_a$, then dividing it by
$\pi^{\frac{3}{2}}\tau_{0}w_{0}^2 I_0
[1-exp(\frac{-2r_a^2}{w_a^2})]$/2. \noindent

\begin{figure}[htbp]
   \includegraphics[width=0.4\textwidth]{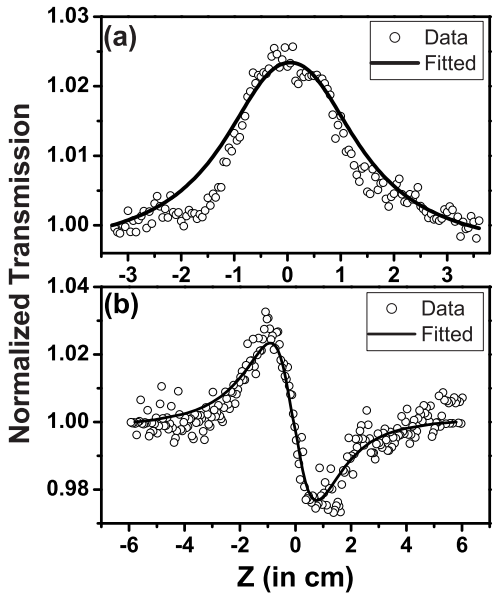}
  \caption{~Normalized transmittance data (open circles) in (a)OA
  z-scan and (b) CA z-scan ($[1-exp(\frac{-2r_a^2}{w_a^2})]$=0.72).
  Theoretical fit (solid line) is obtained with $\beta_{\rm eff} = 1.4~cm/MW$, $I_s= 30~GW/cm^{2}$
  (for OA) and $37~GW/cm^{2}$ (in CA)}
\end{figure}

From the difference between the normalized peak and valley
transmittance in CA z-scan experiment and aperture's linear
transmittance, value of $\gamma$  can be calculated\cite{SB}. From
our measured CA z-scan experiments (shown in Fig. 2(b)), we obtain
a value of $-5.5\times 10^{-11}~cm^2/W$ for $\gamma$. A simple
calculation\cite{note} using the value of $-3\times 10^{-21}~
cm^{-3}$ for ($dn/dN$) predicted\cite{Yu} in wide bandgap
semiconductors also yields a value $-3\times 10^{-11} ~cm^2/W$ for
$\gamma$. However our measured value for $\gamma$ is about two
orders less compared to the value predicted by Margulies et al for
SWNTs\cite{Margulis1}. Using a value of $-5.5\times 10^{-11}~
cm^2/W$ for $\gamma$, we have then varied $\beta_{\rm eff}$ in the
calculation of $T_{CA}$(z) to fit the data. We found that both OA
and CA data can be fitted simultaneously with one set of
parameters as shown by solid line in Fig. 2. With larger values of
$\beta_{\rm eff}$, a good fit to CA data could not be achieved for
any value of $I_s$ (the same is true for OA data too). Although
good fit is possible with much smaller  $\beta_{\rm eff}$, the
$I_s$ required to fit CA differs considerably (by more than an
order of magnitude) from that needed to fit OA data. A consistent
set of parameters for $\gamma$,~$\beta_{\rm eff}$ and $I_s$
respectively, $-5.5\times 10^{-11} ~cm^2/W$, $1.4 \times 10^{-6}~
cm/W$, and $30~GW/cm^2$, fit both OA and CA data well as shown by
solid lines in Fig. 2. As seen from Eq. 3, the $\beta_{\rm eff}$
has two contributions -TPA and FCA. Noting that the FCA
contribution depends on the pulse width, we repeated both CA and
OA z-scan with FWHM width of $140~fs$ to evaluate the relative
strength of these two contributing mechanisms. We found that
z-scan data with $140~fs$ pulse width is identical to that
obtained with $80 fs$ pulse width. This clearly indicates that FCA
cross section is extremely small and the fundamental TPA
($\beta_0$) is the dominant mechanism for nonlinear absorption.
The predicted value of $1.4 cm/MW$ for $\beta_0$ in CNTs is
two-to-three orders of magnitude larger than that in wide band gap
semiconductors at $1$ to $2$ $\mu m$ wavelength. This large value
of nonlinearity (both $\beta_0$ and $\gamma$) arises mainly as the
consequence of the one dimensional motion of delocalized-electron
cloud along the nanotube axis.\cite{Margulis2,Lauret} Although the
large value of $\beta_0$ at the wavelength, where $\alpha_0$ is
also large, makes it less useful for optical limiting
applications, the underlying mechanism responsible for $\beta_0$
could be operative even at the forbidden wavelength. With our
increasing ability to tune the band gap with nanotubes' radius,
CNTs offer an interesting possibility for enhanced nonlinearity in
near visible to short wave infrared wavelength region. \noindent

   AKS thanks Department of Science and Technology, India for financial support and SK thanks the Indian
Institute of Science for his two-month sabbatical visit.

%\newpage


\begin{thebibliography}{10}

\providecommand*{\bibinfo}[2]{#2} \providecommand*{\eprint}[1]{#1}
\providecommand*{\url}[1]{#1}

\bibitem{Chen}
P. Chen, X. Wu, X. Sun, J. Lin, W. Ji, and K. L. Tan, Phys. Rev.
Lett. ~\textbf{82}, 2548 (1999).

\bibitem{Liu}
X. Liu, J.Si, B. Chang, G. Xu, Q. Yang, Z. Pan, S. Xie and P.Ye,
Appl. Phys. Lett. ~\textbf{74}, 164 (2000).

\bibitem{sood}
S.R. Mishra,  S.C Rawat,S.C Mehendale, K.C Rustagi, A.K. Sood, R.
Bandyopadhyay, A. Govindaraj, C.N.R. Rao, Chem. Phys. Lett.
~\textbf{317}, 510(2000).

\bibitem{Vivien}
L. Vivien, D. Riehl, P. Lancon, F. Hache, E. Anglaret, Opt. Lett.
~\textbf{26}, 223(2001).

\bibitem{Sun}
X. Sun, R. Q. Yu, G. Q. Xu, T. S. A. Hor, and W. Ji, Appl. Phys.
Lett. ~\textbf{73}, 3632(1998).


\bibitem{Maeda}
Maeda, S. Matsumoto, H. Kishida,T. Takenobu, Y. Iwasa, M.
Shiraishi,M. Ata and H. Okamoto, \emph{et al.,} Phys. Rev. Lett.
\textbf{94}, 047404 (2005).

\bibitem{achiba}
Y. Sakakibara, S. Tatsuurai,H. Katura, M. Tokumoto and Y. Achiba,
Jpn. J. Appl. Phys. ~\textbf{42}, 494 (2003).

\bibitem{Sheng}
C. -X. Sheng and Z.V. Vardeny, Phys. Rev. Lett. \textbf{96},
019705 (2005).

\bibitem{Okamoto}
H. Okamoto, S. Matsumoto, A. maeda, Y. Kishida, Y. Iwasa, T.
Takenobu, Phys. Rev. Lett, \textbf{96}, 019706 (2006).


\bibitem{Ajayan}
Y.C. Chen, N.R. Raravikar, L.S. Schadler, P. M. Ajayan, Y. P.
Zhao, T.M. Lu, G.C. Wang and X.C. Zhang, Appl. Phys. Lett.
\textbf{81}, 975 (2002).

\bibitem{Tatsuura}
S. Tatsuura, M. Furuki, Y. Sato, I. Iwasa, M. Tian, and H. Mitsu,
Adv. Mater. \textbf{15}, 534 (2003).

\bibitem{SB}
M. Sheik-Bahae, A.A. Said, T.H. Wei, D.J. Hagen, E.W.Van Stryland,
IEEE J.Quant.Electron. \textbf{26}, 760 (1990).

\bibitem{Wearie}
D. Wearie, B. S. Wherrett, D. A. B. Miller and S. D. Smith, Opt.
Lett. \textbf{4}, 331 (1979).

%\bibitem{Kwak}
%C. H. Kwak, Yeung Lak Lee and Seong Gyu Kim, J. Opt. Soc. Am. B,
%\textbf{16}, 600 (1999).

\bibitem{Chapple}
P. B. Chapple, J. Staromlynska, J. A. Hermann and T. J. Mckay, J.
Nonlin. Opt. Phys. and Mater. \textbf{6}, 251 (1997).

%\bibitem{Hermann}
% 20. J. A. Hermann, Int. J. Nonlin. Opt. Phys. \textbf{1}, 541
%(1992).

\bibitem{Yin}
M. Yin, H. P. Li, S. H. Tang, W. Ji, Appl. Phys. B \textbf{70},
587 (2000).

\bibitem{Gao}
Y. Gao, X. Zhang, Y. Li, H. Liu, Y. Wang, Q. Chang, W. Jiao, Y.
Song, Opt. Commun. \textbf{251}, 429 (2005).

%%\bibitem{Pushpa}
%%P.A. Kurian, C. Vijayan, C.S.S. Sandeep, R. Philip and K.
%%Sathiyamoorthy, Nanotechnology \textbf{18},(2007) 075708.
\bibitem{Yang}
L. Yang, R.Dorsinville, Q.Z. Wang, P.X. Ye, R.R. Alfano, R.
Zamboni and C. Taliani, Opt. Lett. \textbf{17}, 323 (1992).

\bibitem{Samoc}
M. Samoc, A. Samoc, B. Luther-Davies, H. Reisch and U. Scherf,
Opt. Lett. \textbf{16}, 1295 (1998).

\bibitem{Gradshteyn}
I. S. Gradshteyn and I. M. Ryzhik, "Table of Integrals, Series,
and Products", (Acadamic Press Inc, California), pp 68-69 (1992).

\bibitem{note}
The $\Delta$n arising from the photo generated carriers is
$[\Delta$N ${\frac{dn}{dN}}]$. Then $\gamma$ is simply
${\frac{dn}{dN}{\frac{\alpha_0\tau_0}{\hbar\omega}}}$

\bibitem{Yu}
Z. G. Yu, S. Krishnamurthy, S. Guha, J. Opt. Soc. Am. B
\textbf{23}, 2356 (2006)

\bibitem{Margulis1}
 27. V1. A. Margulis and E. A. Gaiduk, J. Opt. A:Pure Appl.
Opt. \textbf{3} , 267 (2001)

\bibitem{Margulis2}
Vl. A. Margulis and T. A. Sizikova, Physica B \textbf{245}, 173
(1998).

\bibitem{Lauret}
J-S. Lauret, C. Voisin, G. Cassabois, J. Tignon, C. Delalande, and
Ph. Roussignol, O. Jost, L. Capes, Appl. Phys. Lett, \textbf{85},
3572 (2004).

\end{thebibliography}
\end{document}